\RequirePackage{fix-cm}
\documentclass[twocolumn,epjc3]{svjour3}  

\usepackage{cite}
\usepackage{amsmath,amssymb,amsfonts}
\usepackage{algorithmic}
\usepackage{graphicx}
\usepackage{textcomp}
\usepackage{xcolor}
\usepackage{verbatim}
\usepackage{algorithm}
\usepackage{url}
\usepackage{subcaption}

\smartqed  
\RequirePackage{graphicx}
\begin{document}

\title{Neutralization of IMU-Based GPS Spoofing Detection using external IMU sensor and feedback methodology
}

\author{Ji Hyuk Jung\thanksref{e1,addr1}
        \and
        Ji Won Yoon\thanksref{e2,addr1}
}

\thankstext{e1}{e-mail: graycat@korea.ac.kr}
\thankstext{e2}{e-mail: jiwon\_yoon@korea.ac.kr}



\institute{School of Cybersecurity, Korea University, Seoul 02841, South Korea \label{addr1}
}


\maketitle

\begin{abstract}
Autonomous Vehicles (AVs) refer to systems capable of perceiving their states and moving without human intervention. Among the factors required for autonomous decision-making in mobility, positional awareness of the vehicle itself is the most critical. Accordingly, extensive research has been conducted on defense mechanisms against GPS spoofing attacks, which threaten AVs by disrupting position recognition. Among these, detection methods based on internal IMU sensors are regarded as some of the most effective. In this paper, we propose a spoofing attack system designed to neutralize IMU sensor-based detection. First, we present an attack modeling approach for bypassing such detection. Then, based on EKF sensor fusion, we experimentally analyze both the impact of GPS spoofing values on the internal target system and how our proposed methodology reduces anomaly detection within the target system. To this end, this paper proposes an attack model that performs GPS spoofing by stealing internal dynamic state information using an external IMU sensor, and the experimental results demonstrate that attack values can be injected without being detected.
\end{abstract}

\section{Introduction}
\label{intro}
Autonomous Vehicles (AVs) are defined as systems capable of autonomous operation without human control. In particular, autonomous operation encompasses all domains, including air, land, and sea. They can be deployed in dangerous environments in place of humans, and they are increasingly utilized in sectors such as agriculture and logistics. More recently, their strategic significance within the defense industry has drawn considerable attention. The foundation of autonomous AVs mobility lies in the ability to estimate the dynamic states of the vehicle. Such estimation serves as the basis for autonomous decision-making. Among the various dynamic states of a AVs, positional information is of utmost importance. While research efforts have explored position estimation methods using spatial data from sensors such as Lidar and cameras, the most fundamental sources of positional data remain those derived from Inertial Measurement Units (IMUs) and Global Positioning System (GPS) signals. In particular, GPS remains the only sensor currently capable of providing absolute position information, thus necessitating GPS-based operation for AVs. However, despite its advantages, GPS system also presents vulnerabilities. Since the signals are unidirectionally received from satellites, bidirectional communication is not possible, which leads to inherent security weaknesses. In light of the growing importance of AVs across both defense and commercial industries, substantial research has been directed toward attack methodologies and corresponding countermeasures \cite{kerns2014unmanned, humphreys2008assessing, tippenhauer2011requirements, feng2018efficient, sathaye2022semperfi, tanil2016kalman}. One such threat is GPS spoofing, in which an adversary manipulates satellite signal information by transmitting counterfeit signals with greater strength than authentic satellite signals. As a result, the AV is deceived into recognizing a false position as its actual location. To mitigate such attacks, defense mechanisms have been proposed, including encryption techniques and the utilization of wireless signal strength and directionality\cite{cheng2009authenticity, warner2003gps,meurer2016direction}. External signals can also be accessed by attackers, which makes them limited against sophisticated adversaries. Another approach is to use localization through internal sensors to detect anomalies in GPS signals. Since attackers have difficulty accessing a AV’s internal sensors and this method relies solely on them, it becomes an efficient solution. Detection techniques based on internal sensors are not only secure against attackers but have also evolved into a robust method \cite{jafarnia2012detection, wendel2006integrated, agyapong2021efficient, khanafseh2014gps, feng2018efficient, tanil2016kalman, nayfeh2023machine, jullian2021deep}. In contrast to these advances in detection techniques, attempts to develop attack methods that can bypass them remain limited. Most studies on GPS attacks often address attack success while ignoring detection techniques, and only rarely attempt to maintain attacks while evading detection.
Although some research has been conducted on GPS attacks designed to bypass detection, they often assume theoretically modeled systems where the attacker is granted access to the internal system states \cite{su2016stealthy, kwon2020performance}. Therefore, this paper proposes a more realistic attack model that rigorously aims to neutralize detection and continuously conduct GPS spoofing attacks. Furthermore, we validate the feasibility of this model using PX4, the most widely referenced AV autopilot system in both research and industry \cite{px4}.

Recent studies have suggested that GPS attacks can be carried out under the assumption that an attacker is able to determine the target's position by using high-precision radar equipment, by following the vehicle, or by observing it from a specific location, and they have also proposed corresponding countermeasures. \cite{jung2024analysis, zhang2025ghost}. However, while methodologies and experiments have been presented, there is a limitation in that the use of high-precision radar equipment was only assumed rather than actually implemented. Following a vehicle using LiDAR or ambushing it from a specific location may be feasible for automotives, but it is difficult for aerial AVs such as AVs that have more freedom of movement. In this paper, we propose an approach in which the attacker neutralizes detection by estimating the internal IMU sensor values of the target system through the use of an external IMU sensor. The attack model assumes that if a external attack system is mounted on the AV, the attacker can automatically carry out GPS spoofing while evading AV detection. Notably, this approach can also be applied to various autonomous systems, such as self-driving cars and Urban Air Mobility (UAM). Since IMU sensors are inherently subject to significant bias, even sensors of the same type produce differing measurements \cite{sahawneh2008development}. However, our experiments demonstrate that these differences can be overcome. Finally, internal detection in AVs is based on internal state estimates from the IMU sensor. GPS spoofing by the attacker not only affects these estimates directly but also interacts with changes caused by the internal IMU itself. To minimize this impact, we devised a method in which spoofed values are iteratively reflected in subsequent spoofing signals. In summary, the contributions of this paper are as follows:

\begin{itemize}
    \item We propose an attack model using an external IMU to neutralize IMU-based detection.

    \item We experimentally demonstrate the feasibility of neutralizing detection that relies on internal IMU values through the use of an external IMU.

    \item To enable continuous spoofing, we propose a method in which spoofing values are adjusted by incorporating previously injected spoofing signals in addition to IMU readings, allowing the attacker to compensate for the reflections observed in the target system.
\end{itemize}

\section{Background}
\label{sec:1}

\subsection{GPS Spoofing Attack and Defense}
\label{sec:2}

In AV sensors, IMUs, magnetometers, and vision sensors can be used to estimate relative positions, but GPS sensor is the only means to directly obtain absolute position coordinates \cite{spilker1978gps}. Therefore, unless operating in confined indoor environments, GPS signals are absolutely crucial for AVs. However, GPS reception relies on signals received from satellites, making two-way communication difficult and, by nature, hard to encrypt. This is because a typical GPS receiver cannot transmit signals back to the satellites. Consequently, GPS systems are inevitably highly vulnerable to spoofing attacks. As a result, extensive research has been conducted on attack methods against GPS and corresponding detection techniques. Fundamentally, GPS spoofing deceives a AV by transmitting radio signals stronger than authentic satellite signals, thereby delivering false information \cite{tippenhauer2011requirements}. Hence, the first layer of spoofing detection methods involves leveraging the directionality and characteristics of these radio signals \cite{jafarnia2012detection, montgomery2009receiver, sathaye2022semperfi}. However, this approach can incur additional hardware costs, and since radio signals are external, attackers can also manipulate or adapt to them, posing inherent limitations. Therefore, it has been proposed that GPS spoofing should be detected within the AV’s navigation system itself. Internal systems are more secure than methods relying solely on external signals since they are less accessible to attackers, and they are also more efficient as they can be implemented through software algorithms \cite{jafarnia2012detection, wendel2006integrated, agyapong2021efficient, khanafseh2014gps, feng2018efficient, tanil2016kalman}.

\subsection{EKF Sensor Fusion in AVs}\label{subsec:sensor_fusion}

\begin{figure}[]
\centering
\includegraphics[width=6cm, height=4.12cm]{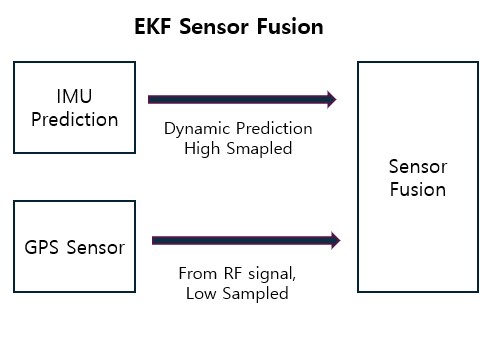}
\caption{Position estimation using sensor fusion with GPS and IMU.\label{fig:sensorFusion}}
\end{figure}

In general, AV’s position estimation systems are primarily based on IMU sensors and GPS sensors. By fusing these two sensors, it becomes possible to estimate the AV’s position with high precision, as they function complementarily. The IMU sensor, with its high sampling rate, can estimate position over short periods of time. However, due to its high drift, errors accumulate as time passes. On the other hand, GPS sensor has a relatively low sampling rate and relies on external satellite signals, which may not always be available. Nevertheless, it provides absolute position information and has the advantage that its position errors do not accumulate over time. Currently, most AVs estimate their position by fusing multiple sensors and utilizing dynamic motion models, with Extended Kalman Filter (EKF) sensor fusion commonly employed for this purpose. EKF sensor fusion is an algorithm that provides optimal state estimation by combining sensor measurements with system modeling \cite{kalman1960new}. Basically, the position estimation process of AVs using EKF sensor fusion can be divided into a prediction process based on the AV's dynamic model and a process using GPS sensor measurements. The dynamic model updates the position change using acceleration and angular velocity measured by the IMU, while the process using measurements incorporates GPS values into the estimated state \cite{sola2017quaternion, feng2018efficient}. Figure \ref{fig:sensorFusion} summarizes the sensor fusion process of GPS and IMU in AVs.

In this section, since the focus is mainly on position estimation, the details on vehicle attitude and velocity estimation are omitted. Additionally, only the estimation using GPS and IMU is explained, excluding other sensors such as magnetometers or additional sensors. IMU sensor measures the current angular velocity and the acceleration along the x, y, and z axes from its basic position. If the sensor measurement sampling interval of IMU is denoted as $\Delta t$, the dynamic state prediction by IMU at each step can be expressed as follows:

\begin{equation}
\begin{aligned}
&x = [p_N, p_E, p_D, v_N, v_E, v_D] \\
&u = [a_N, a_E, a_D] \\
&v_{t+\Delta t} = v_t + a_t \\
&p_{t+\Delta t} = p_t + a_t \times \Delta t + \frac{1}{2} \times a_t \times \Delta t ^ 2. \\
\end{aligned}\label{eq:newton}
\end{equation}
In Equation \ref{eq:newton}, $x$ represents the state of AVs and is expressed in the NED coordinate frame. In this section, only position and velocity are considered for equation development. $u$ denotes the acceleration measured by the IMU, while $v$ and $p$ correspond to the equations underlying the IMU's position prediction process based on Newtonian dynamics. The process of EKF sensor fusion can be summarized as follows:
\begin{equation}
\begin{aligned}
&\hat{x}_{k|k-1} \leftarrow f(\hat{x}_{k-1|k-1}, u_{k}) \\
&P_{k|k-1} = F_kP_{k-1|k-1}F_k^T + Q_k  \\
&S_k = H_kP_{k|k-1}H_k^T + R_k  \\
&K_k = P_{k|k-1}H_k^TS_k^-1 \\
&\hat{x}_{k|k} \leftarrow \hat{x}_{k|k-1} + K_k \times (z_{k} - h(\hat{x}_{k+1}))
\end{aligned}
\end{equation}
where the dynamic model $f$ predicts $x_{k}'$ based on the system's current internal estimates $x_{k-1}$ and the IMU sensor input $u_{k}$. The predicted value mentioned in Equation \ref{eq:newton} is modeled by adding noise to Newtonian dynamics. Here, since Newtonian dynamics is nonlinear and noise is added, linearization is performed using the Jacobian matrix, and the noise is addressed using a Gaussian model. In other words, a Gaussian linear model can be derived, where $F$ is the Jacobian matrix of the dynamic function $f$. After prediction by $F$, the accuracy of the prediction, the covariance, also changes and thus needs to be updated. $P_{k|k-1}$ represents the predicted covariance at step $k$, reflecting information up to step $k-1$. This is updated by multiplying the predicted covariance from step $k-1$ with the linearized Jacobian matrix and adding the system noise $Q$. $Q$ is generally fixed during use. The function $h$ defines the relationship between the sensor measurements and the internal state estimates. $H$ is the Jacobian matrix of $h$, the relationship between the GPS sensor and the state, based on the same principle as $F$. In the sensor measurement update, the update is performed using the difference between the GPS measurement $z$ and the predicted value. At this time, the covariance of the difference with the measurement is calculated and represented by $S$. $S$ is computed by multiplying the predicted covariance by the measurement Jacobian $H$ and then adding the sensor's own noise $R$. Finally, the Kalman Gain, which indicates the extent to which the measurement error is reflected in the predicted state, is calculated using this. Through this, the estimated value $\hat{x}_{k|k}$ at step $k$ is updated. It should be noted, as mentioned in Figure \ref{fig:sensorFusion}, that the measurements provided by the GPS differ in sampling rate and resolution from those of the IMU, meaning they are not synchronized. Therefore, GPS values only update the state at specific time points. Consequently, the measurement update process occurs only when GPS data is sampled, while at times when GPS data is not sampled, the update step is skipped, and only the prediction step is performed.

\subsection{IMU-Based Spoofing Detection}
In this section, we describe GPS spoofing detection methodology based on IMU sensor. The fundamental idea is to detect anomalies by continuously comparing the position changes estimated from IMU sensor with GPS signals received from external sources. Since IMU operates internally within AVs and cannot be accessed by an attacker without internal intrusion, its measurements can serve as the most secure reference. In EKF sensor fusion, innovation is computed each time GPS measurements are incorporated. This innovation is then used to calculate a test ratio, which determines whether the GPS signal is anomalous. Typically, when the test ratio exceeds 1, the GPS measurement is regarded as spoofed or otherwise abnormal\cite{px4, liu2019analysis, quinonez2020savior, noh2019tractor, jung2024analysis}. Innovation, represented by $(z_{k} - h(\hat{x}_{k+1}))$, refers to the discrepancy between the sensor measurement and the predicted internal state estimate. Innovation variance $S_k$ (or covariance), determined by preditcion covariance and the sensor’s inherent sensor noise, represents both the magnitude and the uncertainty associated with the measurement. A larger innovation variance $S_k$ results in a wider allowable range for the innovation, whereas a smaller innovation variance yields a narrower range. This allowable range is quantified by the test ratio $T$, which can be expressed as follows:

\begin{equation}
\begin{aligned}
T = \frac{(z_{k} - h(\hat{x}_{k+1}))^2}{S_k\times G^2}.\\
\end{aligned}
\end{equation}

Innovation Gate, $G$ functions as a threshold to determine the allowable range of the innovation value and is defined as a parameter. Basically, most GPS spoofing detection techniques rely on the difference between sensor-based estimations, especially those centered around IMU sensors, and use various algorithms to detect spoofing \cite{quinonez2020savior, abdo2024avmon, giraldo2018survey}. Recently, many artificial intelligence-based detection methods have also been actively studied \cite{sun2023deep, shafique2021detecting}. Nevertheless, these detection methods ultimately rely on differences from the estimated values. Therefore, in this paper, we conducted experiments using the difference from the most fundamental estimated values as the detection criterion. In this paper, the test ratio value itself is used as a detection criterion to measure the degree of evasion, while, as in \cite{quinonez2020savior}, a method that uses the accumulated sum of difference between the estimated and measured values as a detection criterion for attack detection is also employed.
Since the attacker attempts to injects the position error continuously, detecting such anomalies through the accumulated error is one of the most important processes. Equation \ref{eq:cusum} calculates the accumulated error, where $S_k$ continuously adds the previous step’s error $r_k$.

\begin{equation}\label{eq:cusum}
\begin{aligned}
&r_k = z_{k} - h(\hat{x}_{k+1})\\
&S_k = S_{k-1} + r_{k}.
\end{aligned}
\end{equation}

In an attack, it is important to continuously follow the internal estimated values. Therefore, this paper measures a coefficient that evaluates whether the overall pattern of the attack injection values is similar to that of the internal estimates. This coefficient is used to quantify how well the attack injection values track the internal estimates, and it can also be utilized for attack detection.  
The coefficient $\rho$ can be expressed as follows, 
\begin{equation}
\begin{aligned}
\rho = \frac{\sum_{t=0}^{T}[z_{k}-u_{m}][h(\hat{x}_{k+1})-u_{e}]}{\sqrt{\sum_{t=0}^{T}[z_{k}-u_{m}]^2\sum_{t=0}^{T}{[h(\hat{x}_{k+1})-u_{e}]^2}}}
\end{aligned}
\end{equation}
, where T is the total time interval over which the patterns are compared, $u_{m}$ is the mean of the measured values, and $u_{e}$ is the mean of the internally estimated values.

\section{Proposed Method}

According to subsection \ref{subsec:sensor_fusion}, in order for an attacker to disable IMU-based detection and successfully carry out a GPS spoofing attack, the spoofed values must continuously remain within the bounds of the internal state estimates. This paper first proposes an attacker modeling method in which the attacker attaches GPS spoofing system to the target system in order to neutralize IMU-based detection and manipulate internal localization. Therefore, the GPS spoofing system can hijack the movements of the target system. This means that it captures only the external motion data without accessing the target’s internal system. To neutralize IMU-based detection, it is necessary to be able to estimate the target system’s internal position estimate. In this paper, we propose a feedback-based spoofing methodology to defeat detection. The approach simultaneously estimates how spoofed GPS values are integrated into the target system, performs spoofing based on the estimated position, and reuses the values processed through the same sensor fusion algorithm within the spoofing system in a feedback loop. In addition, this paper conducts experiments that involve injecting velocity values, which are difficult to handle in conventional GPS spoofing attack models. This is because a real GPS sensor provides not only position information but also velocity data. Therefore, if the injected attack values fail to follow the internally estimated velocity, they can also be detected. To evaluate how well the injected velocity follows the internal estimates, this study compares them with actual GPS sensor data. Unlike position spoofing, velocity is determined instantaneously by internal sensors rather than being accumulated over time. Hence, in the velocity injection experiments, the feedback-based method was not applied; instead, values estimated by an external IMU sensor were continuously injected. To evaluate how well the injected values followed the internal estimates, the experiment used both a coefficient to measure pattern similarity and the difference between the internal estimates and injected values, referred to as the innovation. Since the magnitude of innovation increases with the internal estimated velocity, it was normalized by dividing by the internal estimated velocity to obtain a relative comparison.

\subsection{Attack Model}\label{subsec:AttackModel}
In this section, we assume that an attacker can physically attach a GPS spoofing system to AVs, automobile, or other target through external access. It is even more feasible than assuming access to high‑cost radars capable of real‑time position estimation. Moreover, by exploiting this assumption, an attacker can perform GPS spoofing that is even more precise than the spoofing methodology based on location. This is because, for the attacker, the most critical factor is not the target system’s true position but rather its internally estimated position, and methods that exploit the IMU sensor used for internal position estimation can achieve greater precision in tracking that the target's internal position.

\begin{figure}[]
\centering
\begin{subfigure}[b]{7cm}
    \centering
    \includegraphics[width=5.87cm, height=3.46cm]{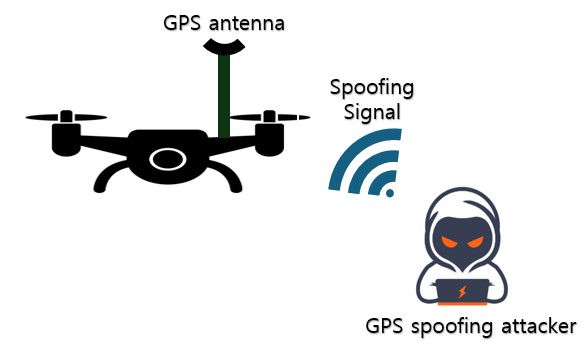}
    \caption{Existing spoofing attack model, where the attacker sends GPS spoofing signals remotely to the AV from a distant location.}
    \label{fig:AttackModel_Existing}
\end{subfigure}
\begin{subfigure}[b]{7cm}
    \centering
    \includegraphics[width=6.18cm, height=4.96cm]{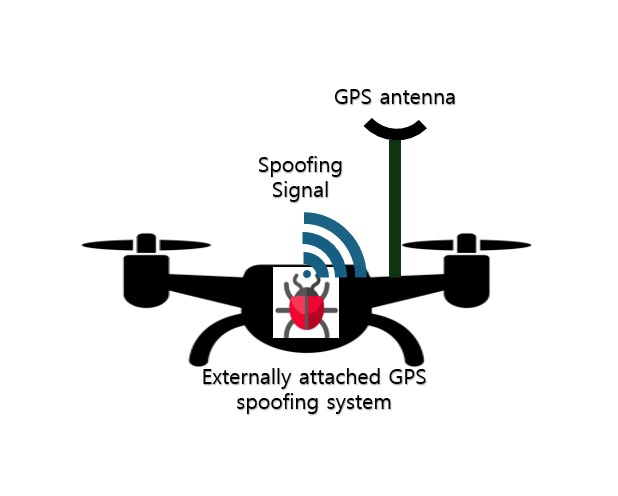}
    \caption{Proposed spoofing attack model in this paper, where the attacker attaches a external spoofing system to the AV and conducts GPS spoofing attacks.
}
    \label{fig:AttackModel_Proposed}
\end{subfigure}
\vspace{0.5cm}

\caption{Comparison between the proposed GPS spoofing attack model in this paper, which attaches a external spoofing system to AVs for direct spoofing, and the existing attack model where the attacker sends spoofing signals remotely from a distance.
}
\label{fig:AttackModel}
\end{figure}

\subsection{Neutralization Methodology of Spoofing Detection}\label{subsec:NeuMethod}

\begin{figure*}[!t]
\centering
\includegraphics[width=12cm, height=5.3cm]{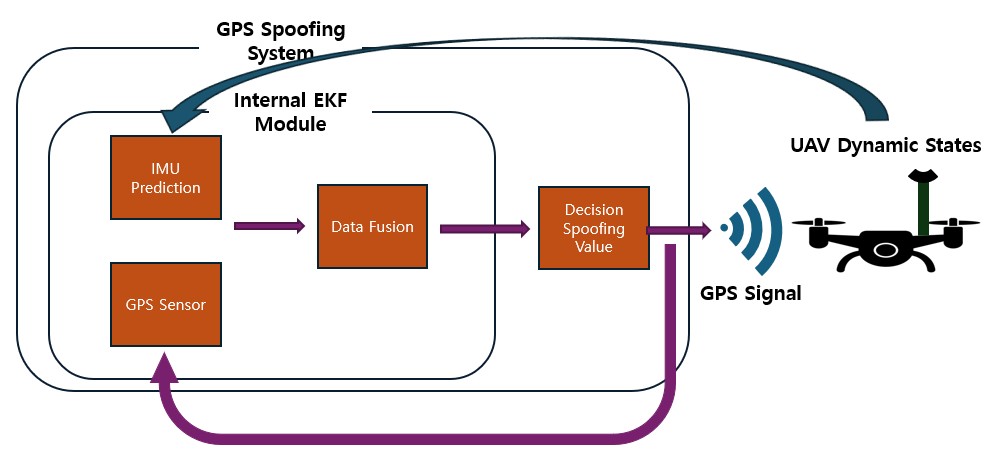}
\caption{GPS spoofing system for Neutralization IMU-based Detection.}
\label{fig:SpoofingSystem}
\end{figure*} 

For an attacker to carry out an attack without being detected, the spoofing signal must consistently remain within the range of the target system’s internally estimated position. Although this requirement can be highly challenging for the attacker, one key principle can be exploited. The estimated position of the target system gradually converges toward the value induced by the attacker’s GPS spoofing signal. In other words, while the attacker needs to estimate the target’s position, the attacker simultaneously has the ability to influence, to some degree, the very position that must be estimated. Once the GPS spoofing signal is reflected in the target system’s position estimation, the internal estimate shifts closer to the injected spoofed value by a certain margin. This principle enables continuous spoofing attacks even if discrepancies from internal position estimation accumulate during intervals when the GPS spoofing is not being applied. However, if the rate at which the target system’s position diverges during intervals where GPS spoofing is not applied becomes too rapid, it becomes increasingly difficult for the attacker to accurately predict the target’s position using spoofing alone. In such cases, a methodology to compensate for this positional drift is required. The most straightforward approach would be to assume that the attacker has access to a radar system or knowledge of the target system’s actual position. However, this assumption has the drawback of requiring prohibitively expensive equipment.

In this paper, we propose an alternative methodology that compensates for positional drift by leveraging an external IMU sensor to estimate variations in the target system’s IMU data, and incorporating this estimate into the spoofing attack. This approach presents a realistic threat to AVs and autonomous vehicles in practice. Figure \ref{fig:SpoofingSystem} illustrates the GPS spoofing methodology proposed in this paper, while Algorithm \ref{alg:spoofing} provides a brief overview of the attack process. 
The AV's mission begins, and the attack system becomes ready to launch the attack at any time. The attack starting point $z_{0}$ is set using the current GPS reception value $z_{gps}$ of the attack system. Once the attack starts, the attack system continuously transmits the system's attack value $z$. This attack value $z$ is predicted based on the AV’s dynamic state estimated through its own IMU to neutralize AV detection, and this prediction is reflected accordingly. Subsequently, the previous attack value is incorporated through sensor fusion. Additionally, the attacker sends the attack value by adding the intended position change $p_{t+1}$ to it. To neutralize IMU-based detection, the proposed methodology was designed by considering two aspects. The first approach is that, in order to compensate for the accumulated positional changes caused by AV dynamics, the attack system uses its own IMU system to predict and correct these variations. Subsequently, when the attack values are applied to the AV, as mentioned in subsection \ref{subsec:NeuMethod}, the Kalman gain is calculated, and the attack values are incorporated accordingly. At this point, the attacker must also reflect this adjustment in the next step of the attack values to ensure that the injected values can track the AV’s internal state estimates. To achieve this, the attack values are also updated and incorporated within the attacker’s own system.

\begin{algorithm}
\caption{Proposed Spoofing System}\label{alg:spoofing}
\begin{algorithmic}
\STATE $Spoofing \, Attack \, Start$
\STATE $t \gets 0$
\STATE $z_{0} \gets z_{gps}$
\WHILE{$!done$}
\STATE $predict \, x_{t+1} \, based \, on \, AV's \, dynamic \, state$
\STATE $fuse \, z_{t} \, with \, x_{t+1}$
\STATE $z_{t+1} \gets x_{t+1} + \Delta p_{t+1}$
\STATE $transmit \, z_{t+1}$
\STATE $t \gets t + 1$
\ENDWHILE
\end{algorithmic}
\end{algorithm}

Additionally, this paper considers the detection of velocity values as well as position values. Of course, the objective of velocity spoofing is to provide values that do not cause a significant change in the internal estimation, and it is not to change the internal value itself. That is, the injected value focuses not merely on changing the internal value, but on whether it can follow the internal change. Furthermore, velocity is directly affected by changes in the IMU sensor values, leading to a very large fluctuation range. Therefore, for the velocity value, the velocity estimated by the attacker is directly injected without going through feedback, minimizing the changes to the attacker's value or the internal value. In other words, the goal is not to allow the attacker to control the value through continuous value injection, but rather to reduce the detection by AV by injecting values that do not significantly deviate from the internal values.

\section{Experiment}
The experiments in this study focus on attacks and detection conducted within the AV control system. Therefore, experimentally spoofing actual GPS signals is prohibited by national regulations, and generating precise GPS signals using expensive equipment can be burdensome for other experiments. To design a spoofing experiment that closely resembles realistic conditions, this paper utilizes MAVLink radio signals to transmit GPS signals.

\begin{figure*}[!t]
\centering
\includegraphics[width=12cm, height=4.8cm]{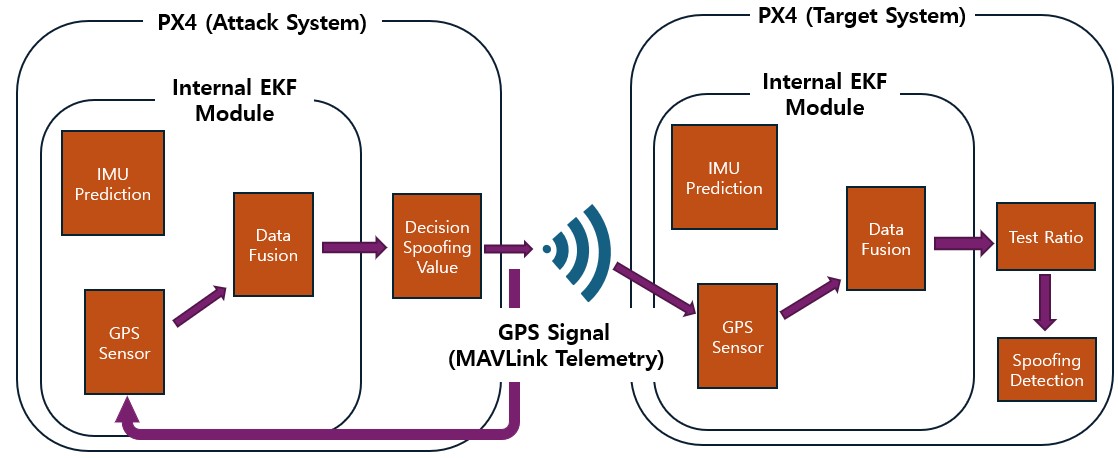}
\caption{Experimental environment using MAVLink with GPS spoofing.\label{fig:Exp}}
\end{figure*}

\subsection{Implementation Details}

Although transmitting spoofed GPS signals through MAVLink is not natively supported in PX4, we modified MAVLink packets so that the AV perceives them as GPS signals. The controller used in this experiment is a Pixhawk 2.4.8 controller operating on PX4 autopilot platform. Pixhawk 2.4.8 uses an MPU6000 IMU sensor. Additionally, to transmit MAVLink data via radio signals between two controllers, 915 MHz Telemetry SIK radio was used. The objective of the experiment is to determine whether spoofed values, transmitted using an external IMU sensor, can bypass anomaly detection mechanisms within the target system. To achieve this, the experimental setup involves two PX4 controllers: one functioning as the attack system and the other as the target system. The attack system is configured to move in synchronization with the target system, and its IMU sensor is used to attempt to evade detection of the target system's internal state estimates while transmitting spoofed GPS signals.

To send GPS values from the attack system, other MAVLink packets were prevented from being sent. Since the attack system also used a Pixhawk controller, it was observed that default heartbeats and various other packets caused jamming effects in telemetry and MAVLink receiver, which hindered smooth data transmission. Additionally, in this paper, a spoofing module was designed inside the Pixhawk itself to allow spoofing values to be sent without delay. This was because, through experiments using an intermediate module, communication buffer delays were found in the intermediate module, which could affect the experiments. In other words, by integrating the spoofing module directly within the attack system's firmware, immediate spoofing system experiments could be conducted. Figure \ref{fig:Exp} illustrates the approximate experimental setup. The experiment proceeds in three stages. First, we observe the degree of detection when spoofed signals are transmitted without using any external information. Next, we evaluate performance when only positional information from the attacker’s IMU is used. Finally, we analyze the effectiveness when both the IMU’s positional information and feedback from the target system’s response to spoofed values are incorporated in generating subsequent spoofing values. In this subsection, we investigated how test ratio changes when the target system undergoes significant accumulated movement while the attack system remains stationary and sends only a fixed spoofing value. In the velocity injection experiment, we collected data to compare velocity injection from standard GPS sensors(M8N GPS sensor) against IMU-based injection, tracking real GPS sensor responses to IMU velocity changes. These data enabled comparison with results from the proposed external IMU sensor.

\subsection{Basic Methodology}\label{exp:Basic}
\begin{figure}
\centering
\vspace{0.5cm}

\begin{subfigure}[a]{7cm}
    \centering
    \includegraphics[width=6.5cm, height=4.1cm]{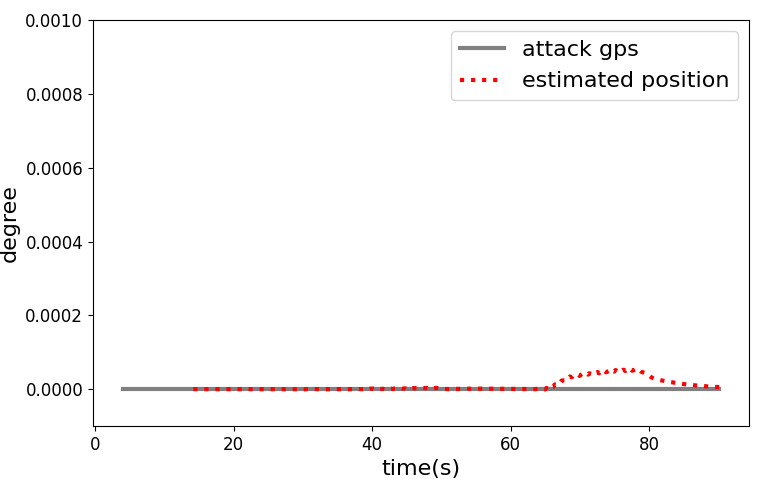}
    \caption{The target system's test ratio.}
    \label{fig:0_acc}
\end{subfigure}

\begin{subfigure}[b]{7cm}
    \centering
    \includegraphics[width=6.5cm, height=3.9cm]{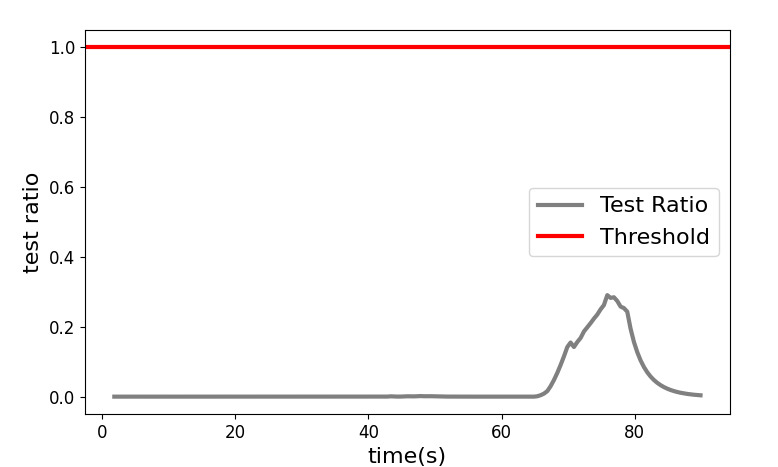}
    \caption{The target system's test ratio.}
    \label{fig:0_acc}
\end{subfigure}

\caption{When the attack system transmits only a fixed spoofing value, changes of the target system's test ratio and the cumulated sum of test ratio.
}
\label{fig:base_exp}
\end{figure}

Under the assumption that the attacker cannot obtain information about the target system’s movements, this experiment examines how the test ratio changes when rapid movement variations are accumulated in the target controller, while the attacker continues to send a constant signal. As shown in Figure \ref{fig:base_exp}, when the signal is transmitted, the magnitude of the difference between the signals sent by the target system and the attacker begins to widen from 60 seconds as the movement variations accumulate. However, at moments when no movement is accumulated, the signals converge again toward the attacker’s constant transmission.

\subsection{Spoofing Methodology Exploiting IMU}\label{exp:IMU}
\begin{figure}
\centering
\begin{subfigure}[a]{7cm}
    \centering
    \includegraphics[width=6.5cm, height=4.1cm]{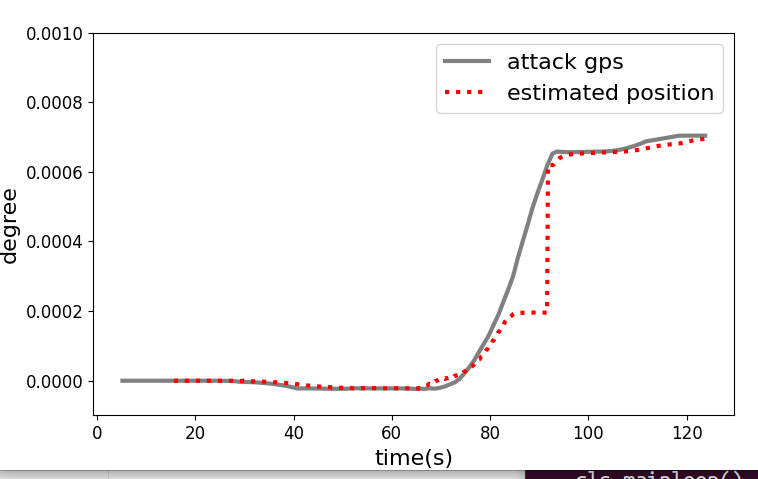}
    \caption{The target system's latitude change (red) and the GPS latitude value sent by the attack system (gray).}
    \label{fig:imu_exp_lat}
\end{subfigure}

\vspace{0.5cm}
\begin{subfigure}[b]{7cm}
    \centering
    \includegraphics[width=6.5cm, height=3.9cm]{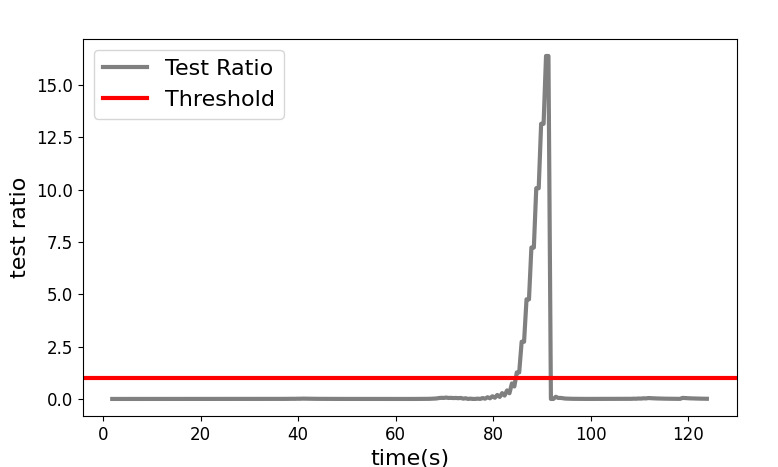}
    \caption{The target system's test ratio.}
    \label{fig:imu_exp_test}
\end{subfigure}
\caption{When the attack system continuously injects values estimated based on IMU sensor into the target system, changes of the target system's latitude and the variation of the anomaly detection value.
}
\label{fig:imu_exp}

\end{figure}

In this experiment, the results are shown when the attack system continuously injects values estimated based on IMU sensor into the target system. In this case, the attack system does not separately reflect the signals it sends, and instead, the target system only incorporates the values estimated solely from IMU without using GPS sensor fusion. As shown in Figure \ref{fig:imu_exp_lat}, the overall movement of the target system is followed, but eventually, due to the error accumulation of the IMU sensor, the estimated values diverge. As this difference accumulates, it ultimately becomes unavoidable for detection in the test ratio.

\subsection{Proposed Feedback Spoofing Methodology}\label{exp:feedback}

Finally, the results obtained using the feedback methodology proposed in this paper are presented. It can be observed that even when rapid movements accumulate, the attack values remain almost within the latitude of the internal estimation. In particular, the test ratio does not exceed 0.014, and by leveraging this, the test ratio can be maintained below 0.5 even when a large attack error value is added.

\begin{figure}
\centering
\begin{subfigure}[a]{7cm}
    \centering
    \includegraphics[width=6.5cm, height=4.1cm]{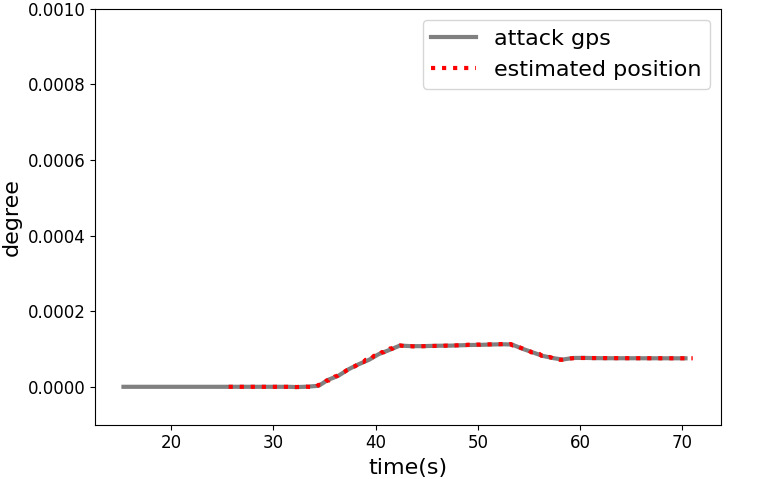}
    \caption{The target system's latitude change (red) and the GPS latitude value sent by the attack system (gray).}
    \label{fig:feedback_exp_lat}
\end{subfigure}

\vspace{0.5cm}
\begin{subfigure}[b]{7cm}
    \centering
    \includegraphics[width=6.5cm, height=3.9cm]{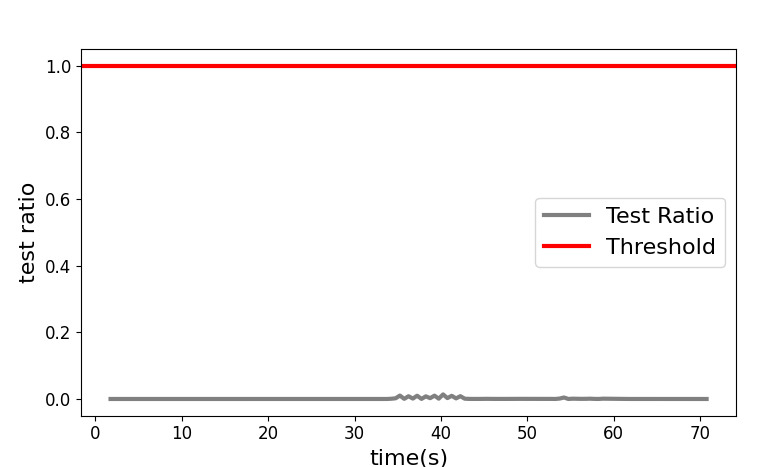}
    \caption{The target system's test ratio.}
    \label{fig:feedback_exp_test}
\end{subfigure}
\caption{when the attack system continuously injects values estimated based on our feedback methodology, changes of the target system's latitude and the variation of the anomaly detection value.
}
\label{fig:feedback_exp}
\end{figure}

\begin{figure}[]
    \centering
    \includegraphics[width=6.5cm, height=3.84cm]{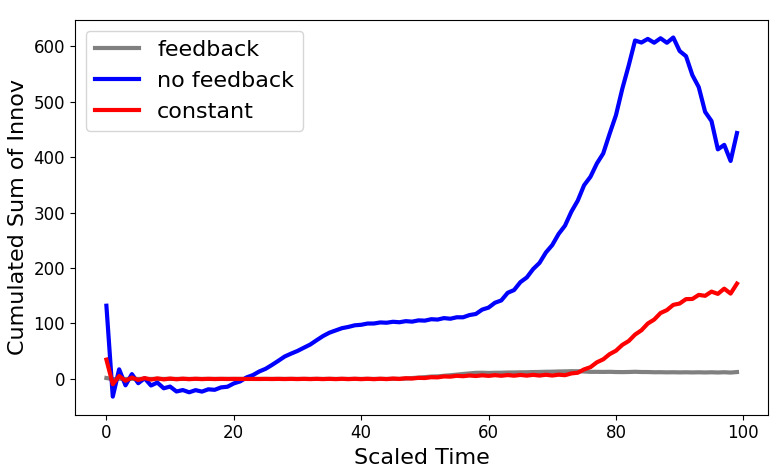}
    \caption{Comparison of the accumulated differences between the estimated and injected values over time for each experiment: the blue line represents using only the IMU without feedback, the red line indicates sending a fixed value, and the last one shows the result using the feedback method.
}
    \label{fig:exp_cusum}
\end{figure}

In this experiment, in addition to test ratio, we compared the cumulative sum of errors between measured and estimated values over time, and calculated the coefficient values of the measured and estimated values during attack injection periods to assess the degree of pattern matching between them. Figure \ref{fig:exp_cusum} shows the cumulative sum values for each method. The red line represents results when a constant value was injected, the blue line shows results from estimation and injection using IMU only, and the gray line indicates results using the feedback method.

In the case of constant value injection, the coefficient cannot be calculated, and thus the pattern matching degree can be considered nearly nonexistent. The IMU-only estimation and injection method showed a pattern match of approximately 0.967, while the feedback method achieved 0.99. (1 indicates complete pattern matching.)

\subsection{Spoofing Velocity}\label{exp:velocity}

\begin{figure}
\centering
\begin{subfigure}[a]{7cm}
    \centering
    \includegraphics[width=6.5cm, height=4.1cm]{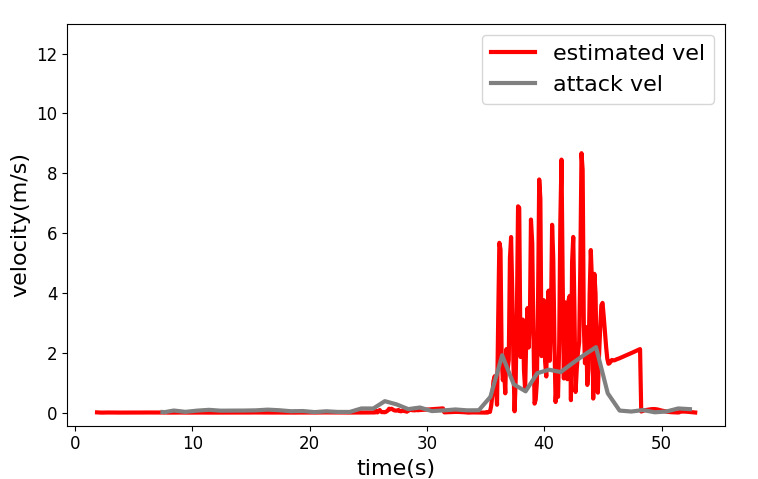}
    \caption{The estimated velocity from GPS M8N sensor (gray line) and the velocity estimated by the internal controller (red line) are shown.}
    \label{fig:vel_m8n}
\end{subfigure}

\vspace{0.5cm}
\begin{subfigure}[b]{7cm}
    \centering
    \includegraphics[width=6.5cm, height=4.1cm]{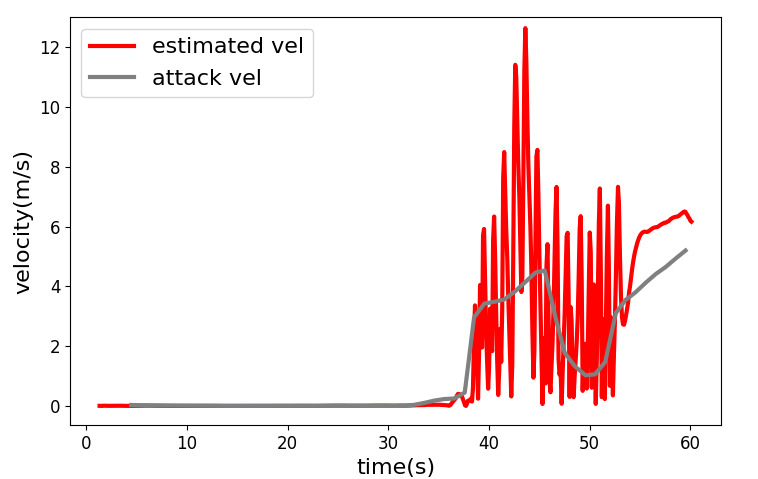}
    \caption{The estimated velocity from IMU sensor (gray line) and the velocity estimated by the internal controller (red line) are shown.}
    \label{fig:vel_imu}
\end{subfigure}
\caption{When estimating internal velocity using GPS, the estimated velocity from the GPS, the velocity estimated by the control system, and the results obtained using an external IMU sensor were compared.
}
\label{fig:exp_vel}
\end{figure}

This experiment aimed to demonstrate that an external IMU can track internal velocity estimates during GPS attacks. It compared the velocity estimated by the actual M8N sensor against estimates from the internal IMU and those using the external IMU sensor, showing that the external IMU can continuously and more accurately follow the speed than GPS estimation alone. Figure \ref{fig:exp_vel} presents the results of this experiment.
The red line represents the internal estimated value, while the gray line represents the attack injection value based on actual GPS values and IMU estimation.

\begin{figure}[]
    \centering
    \includegraphics[width=6.5cm, height=3.93cm]{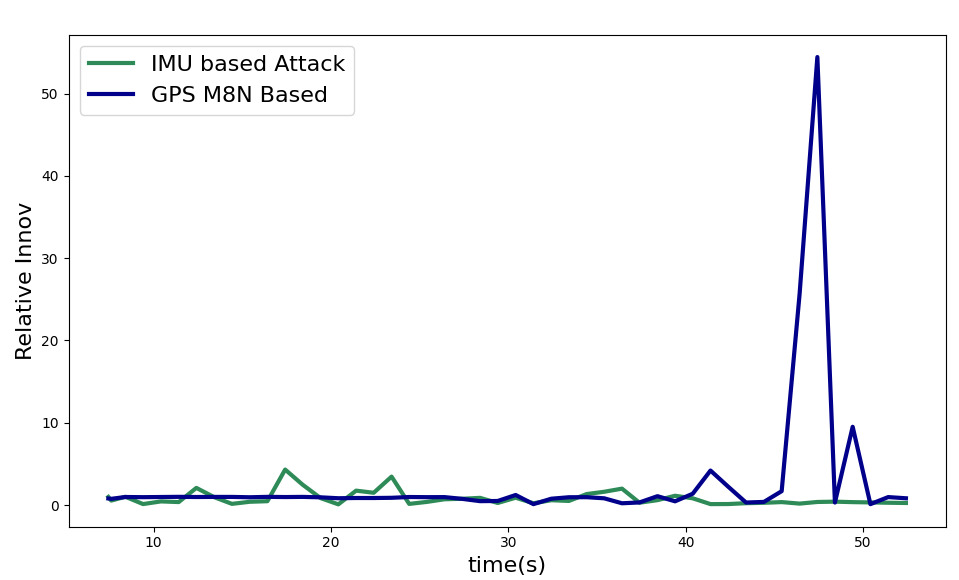}
    \caption{GPS internal velocity estimation error by GPS (blue line) and by external IMU (green line).}
    \label{fig:exp_vel_innov}
\end{figure}

Figure \ref{fig:exp_vel_innov} displays the relative errors against the internal estimates over time, clearly revealing the superior estimation accuracy of the IMU. The blue line represents the error from the GPS sensor, and the green line represents the error from IMU estimation. To assess how well the internal velocity patterns are followed, coefficients were measured: 0.76 for GPS M8N and 0.87 for IMU-based estimation, confirming that IMU usage better captures velocity patterns.

\section{Discussion}
The experimental goal of this paper was to conduct tests on whether it is possible to inject anomalies by bypassing GPS anomaly detection based on IMU in PX4, an autopilot system commonly used in AVs such as AVs and UGVs. In the position injection experiment, the degree of detection avoidance was compared using three detection criteria. The first was the Test Ratio, which represents the instantaneous difference between the internal estimated value and the attack value at each time point. The second was the cumulative sum, which indicates the degree of accumulated difference over a specific period of time. Lastly, the comparison also used Coefficient, which reflects the degree of pattern similarity between the displacement of the internal estimated position and the injected values over a given period.

Subsection \ref{exp:Basic} presents the results when the target system’s movements were accumulated while the attack values continuously spoofed a fixed latitude. The results showed that the test ratio increased. However, once the motion accumulation stopped, the estimation values tended to converge back toward 0 degrees. This result indicates that even if the attacking system does not know the internal estimation values of the target, the target's internal estimations are continuously influenced by the spoofed values through sensor fusion. Moreover, the greater the difference between the spoofed values and the internal estimations, the stronger the convergence effect becomes. In conclusion, even when a constant value is simply injected, the injected value still influences the internal estimation; therefore, the test ratio does not exceed the detection threshold of 1. However, it can be observed that the value is larger compared to the feedback-based methodology proposed in this paper, indicating that detection would be possible if the detection threshold were lowered. The cumulative sum results also showed significantly large values, suggesting that detection could occur as well. Lastly, since a constant value cannot follow the positional pattern of the internal estimations, it can be immediately detected by the third detection criterion. 

Subsection \ref{exp:IMU} shows the results when spoofed values were generated by continuously transmitting position estimates derived from the attacker’s IMU sensor measurements of its own dynamic mobility. In other words, the attack values were set as the attacker’s own internally estimated positions. The experimental results showed that, in broad movement trends, the spoofing values and the target system’s estimates moved together initially, but over time, the difference between the spoofed estimate and the target estimate grew continuously, with the test ratio eventually exceeding 1 by a large margin. This occurs because IMU sensors inherently have significant noise, and position estimation requires double integration, leading to drift that increases substantially over time. Additionally, when the attacker’s spoofed values are fused into the target system via sensor fusion, the target’s estimates are affected, but the attacker does not feed this reflected value back into its next spoofing input. This mismatch further increases the discrepancy. However, unlike attacks that inject constant values, a high coefficient value indicates that the position pattern is well followed. Therefore, this paper proposes a feedback system in which the attacker estimates the target’s movement using its own IMU sensor, while also integrating the spoofed values that were injected into the target back into its own subsequent spoofing calculations. Experimental results using this method, as presented in subsection \ref{exp:IMU}, show that the test ratio was kept very low and that the spoofing attack could be performed in a much more stable manner. In particular, it shows a significant improvement in efficiency compared to the two previous methods. In other words, it shows results with a very small difference from the internal estimates while exhibiting a very high degree of position pattern similarity. The experimental results of this paper indicate that while an attacker can follow the overall movement of the target system using an IMU and perform spoofing attacks, for more precise attacks and stability, it is necessary to reflect the effect of the spoofed values being integrated into GPS sensor fusion. In the velocity injection experiment, the comparison with the injection of actual GPS sensor values demonstrated that velocity values could also be injected without being detected. In other words, the injected values showed smaller differences from the innovation values than those estimated by the real GPS sensor, and the results indicated that the pattern consistency was even higher than that of the GPS sensor.

\section{Conclusion}
GPS attacks have emerged as a serious security threat to drones and autonomous vehicles. However, encrypting GPS signals is nearly impossible due to the inherent limitations of its structure. Therefore, detection algorithms based on IMUs, which are efficient and resistant to internal access such as hacking attempts, have been developed and studied. In this paper, we propose an attack modeling approach that does not assume prior knowledge of location information, but instead considers spoofing attacks executed by an attack system attached to the unmanned vehicle. The attacker’s system, when mounted on the vehicle, can extract the internal IMU sensor data using its own IMU. Next, we analyze the level of detectability when using such IMU data through a test ratio, thereby examining both the effects of GPS sensor fusion on attack data and the consequences of exploiting IMU signals in the attack process. Previously, it was assumed that an attacker could bypass detection or achieve precise spoofing attacks by employing high-performance radar equipment to estimate the target system’s location. Yet, such models fell short of addressing practical considerations such as low-cost implementation, accuracy of location estimation, and the real-time application of positional changes to attack data. In contrast, this paper demonstrates that by applying a feedback system, it is possible to suppress the test ratio significantly and thereby defeat IMU-based detection. Additionally, in this paper, a methodology for experimenting with GPS spoofing using telemetry SiK radio to transmit and receive GPS signals is presented. This will be the best method to transmit GPS values to the controller without expensive equipment during experiments conducted at the controller layer rather than the RF layer in GPS spoofing. To this end, this paper takes a technical approach to customizing MAVLink on the controller and restricts other MAVLink transmissions to avoid radio interference. This will be a good methodology for future research on GPS spoofing at the controller level. This research has important implications for future security studies, as the demonstrated attack method could also be applied to systems such as autonomous cars, drones, and AVs. In future work, research will need to focus on how these attack systems can be effectively defended against.



\begin{thebibliography}{10}
\providecommand{\url}[1]{{#1}}
\providecommand{\urlprefix}{URL }
\expandafter\ifx\csname urlstyle\endcsname\relax
  \providecommand{\doi}[1]{DOI \discretionary{}{}{}#1}\else
  \providecommand{\doi}{DOI \discretionary{}{}{}\begingroup \urlstyle{rm}\Url}\fi

\bibitem{kerns2014unmanned}
A.J. Kerns, D.P. Shepard, J.A. Bhatti, T.E. Humphreys, Journal of field robotics \textbf{31}(4), 617 (2014)

\bibitem{humphreys2008assessing}
T.E. Humphreys, B.M. Ledvina, M.L. Psiaki, B.W. O'Hanlon, P.M. Kintner, et~al., in \emph{Proceedings of the 21st International technical meeting of the satellite division of the institute of navigation (ION GNSS 2008)} (2008), pp. 2314--2325

\bibitem{tippenhauer2011requirements}
N.O. Tippenhauer, C.~P{\"o}pper, K.B. Rasmussen, S.~Capkun, in \emph{Proceedings of the 18th ACM conference on Computer and communications security} (2011), pp. 75--86

\bibitem{feng2018efficient}
Z.~Feng, N.~Guan, M.~Lv, W.~Liu, Q.~Deng, X.~Liu, W.~Yi, ACM Transactions on Embedded Computing Systems (TECS) \textbf{17}(6), 1 (2018)

\bibitem{sathaye2022semperfi}
H.~Sathaye, G.~LaMountain, P.~Closas, A.~Ranganathan, in \emph{Network and Distributed Systems Security (NDSS) Symposium 2022} (2022)

\bibitem{tanil2016kalman}
{\c{C}}.~Tan{\i}l, S.~Khanafseh, M.~Joerger, B.~Pervan, in \emph{2016 IEEE/ION Position, Location and Navigation Symposium (PLANS)} (IEEE, 2016), pp. 1027--1034

\bibitem{cheng2009authenticity}
X.j. Cheng, J.n. Xu, K.j. Cao, J.~Wang, in \emph{2009 Fourth International Conference on Computer Sciences and Convergence Information Technology} (IEEE, 2009), pp. 345--352

\bibitem{warner2003gps}
J.S. Warner, R.G. Johnston, Homeland Security Journal \textbf{25}(2), 19 (2003)

\bibitem{meurer2016direction}
M.~Meurer, A.~Konovaltsev, M.~Appel, M.~Cuntz,   (2016)

\bibitem{jafarnia2012detection}
A.~Jafarnia-Jahromi, T.~Lin, A.~Broumandan, J.~Nielsen, G.~Lachapelle, in \emph{Proceedings of the 2012 international technical meeting of the institute of navigation} (2012), pp. 790--800

\bibitem{wendel2006integrated}
J.~Wendel, O.~Meister, C.~Schlaile, G.F. Trommer, Aerospace science and technology \textbf{10}(6), 527 (2006)

\bibitem{agyapong2021efficient}
R.A. Agyapong, M.~Nabil, A.R. Nuhu, M.I. Rasul, A.~Homaifar, in \emph{2021 IEEE symposium series on computational intelligence (SSCI)} (IEEE, 2021), pp. 01--08

\bibitem{khanafseh2014gps}
S.~Khanafseh, N.~Roshan, S.~Langel, F.C. Chan, M.~Joerger, B.~Pervan, in \emph{2014 IEEE/ION Position, Location and Navigation Symposium-PLANS 2014} (IEEE, 2014), pp. 1232--1239

\bibitem{nayfeh2023machine}
M.~Nayfeh, Y.~Li, K.~Al~Shamaileh, V.~Devabhaktuni, N.~Kaabouch, Computers \& Security \textbf{126}, 103085 (2023)

\bibitem{jullian2021deep}
O.~Jullian, B.~Otero, M.~Stojilovi{\'c}, J.J. Costa, J.~Verd{\'u}, M.A. Pajuelo, in \emph{International Conference on Machine Learning, Optimization, and Data Science} (Springer, 2021), pp. 527--540

\bibitem{su2016stealthy}
J.~Su, J.~He, P.~Cheng, J.~Chen, IFAC-PapersOnLine \textbf{49}(22), 291 (2016)

\bibitem{kwon2020performance}
K.C. Kwon, D.S. Shim, Sensors \textbf{20}(4), 954 (2020)

\bibitem{px4}
P.D. Team.
\newblock Px4 autopilot.
\newblock \url{https://docs.px4.io} (2024).
\newblock Accessed: 2024-10-01

\bibitem{jung2024analysis}
J.H. Jung, M.Y. Hong, H.~Choi, J.W. Yoon, IEEE Access  (2024)

\bibitem{zhang2025ghost}
J.~Zhang, S.~Cheng, L.~Hu, J.~Zhang, C.~Shi, X.~Han, T.~Zhang, Y.~Cheng, W.~Zhang, in \emph{34th USENIX Security Symposium (USENIX Security 25)} (2025), pp. 3979--3998

\bibitem{sahawneh2008development}
L.~Sahawneh, M.~Jarrah, in \emph{2008 5th International Symposium on Mechatronics and Its Applications} (IEEE, 2008), pp. 1--9

\bibitem{spilker1978gps}
J.J. Spilker~Jr, Navigation \textbf{25}(2), 121 (1978)

\bibitem{montgomery2009receiver}
P.Y. Montgomery, T.E. Humphreys, B.M. Ledvina, in \emph{Proceedings of the 2009 international technical meeting of the institute of navigation} (2009), pp. 124--130

\bibitem{kalman1960new}
R.E. Kalman,   (1960)

\bibitem{sola2017quaternion}
J.~Sola, arXiv preprint arXiv:1711.02508  (2017)

\bibitem{liu2019analysis}
Y.~Liu, S.~Li, Q.~Fu, Z.~Liu, Q.~Zhou, IEEE Sensors Journal \textbf{19}(13), 5167 (2019)

\bibitem{quinonez2020savior}
R.~Quinonez, J.~Giraldo, L.~Salazar, E.~Bauman, A.~Cardenas, Z.~Lin, in \emph{29th USENIX security symposium (USENIX Security 20)} (2020), pp. 895--912

\bibitem{noh2019tractor}
J.~Noh, Y.~Kwon, Y.~Son, H.~Shin, D.~Kim, J.~Choi, Y.~Kim, ACM Transactions on Privacy and Security (TOPS) \textbf{22}(2), 1 (2019)

\bibitem{abdo2024avmon}
A.~Abdo, S.M.B. Malek, X.~Zhao, N.~Abu-Ghazaleh, in \emph{Symposium on Vehicle Security and Privacy (VehicleSec)} (2024)

\bibitem{giraldo2018survey}
J.~Giraldo, D.~Urbina, A.~Cardenas, J.~Valente, M.~Faisal, J.~Ruths, N.O. Tippenhauer, H.~Sandberg, R.~Candell, ACM Computing Surveys (CSUR) \textbf{51}(4), 1 (2018)

\bibitem{sun2023deep}
Y.~Sun, M.~Yu, L.~Wang, T.~Li, M.~Dong, Drones \textbf{7}(6), 370 (2023)

\bibitem{shafique2021detecting}
A.~Shafique, A.~Mehmood, M.~Elhadef, IEEE access \textbf{9}, 93803 (2021)

\end{thebibliography}


\end{document}